Article

# Remote State Preparation of Mental Information: A Theoretical Model and a Summary of Experimental Evidence

**Patrizio E. Tressoldi\* and Andrei Khrennikov†**

**ABSTRACT**

The aim of this paper is to define in theoretical terms and summarise the available experimental evidence that physical and mental "objects", if considered "information units", may present similar classical and quantum models of communication beyond their specific characteristics. Starting with the Remote State Preparation protocol, a variant of the Teleportation protocol, for which formal models and experimental evidence are already available in quantum mechanics, we outline a formal model applied to mental information we defined Remote State Preparation of Mental Information (RSPMI), and we summarise the experimental evidence supporting the feasibility of a RSPMI protocol. The available experimental evidence offers strong support to the possibility of real communication at distance of mental information promoting the integration between disciplines that have as their object of knowledge different aspects of reality, both physical and the mental, leading to a significant paradigm shift in cognitive and information science.

NeuroQuantology 2012; 3: ● ● - ● ●

## Introduction

### 1.1 Theoretical background

Is it possible to assume that mental and physical information obeys to common laws? For example, is it possible that phenomena and experimental protocols typical of quantum mechanics can also be observed and applied to mental phenomena? What may be the feature that allows physical and mental "objects" so apparently different, to demonstrate similar "behaviour"?

This is a very old problem. In particular, we remind about a long series of discussions between Pauli and Jung (2001). Pauli thought that the probabilistic nature of quantum theory and the Uncertainty Principle offered the possibility of discovering something beyond the mind-matter gap: "*we must postulate a cosmic order of nature beyond our control to which both the outward material objects and the inward images are subject.*" There is, he thought, a quantum explanation for synchronistic occurrences which somehow "acausally weaves meaning into the fabric of nature." Exploration of this might lead to an answer to the conundrum posed by quantum indeterminacy: if the deepest structures of reality are probabilities then "what fixes what actually happens?" Jung and Pauli sought a unifying theory that would allow interpretation of reality as a psycho-physical whole. Pauli thought that probability mathematics expresses physically what is manifested psychologically as archetypes (deep-structure patterns for certain types of

---
Corresponding author: Patrizio Tressoldi
Address: *Dipartimento di Psicologia Generale, Università di Padova, Italy. †International Center for Mathematical Modelling in Physics and Cognitive Sciences Linnaeus University, Sweden.
Phone: + 390498276623
✉ patrizio.tressoldi@unipd.it

universal mental experience, or patterns of the instincts) and synchronistic events.

The aim of this paper is to define in theoretical terms this possibility by considering physical and mental "objects", as "information units" beyond their specific characteristics. Unifying physical and mental objects in terms of units of information, it is possible to formalise models of transmission of information conveyed by "objects" regardless of their physical or mental nature.

Starting with the Remote State Preparation protocol, a variant of the Teleportation protocol, for which formal models and experimental evidence are already available in quantum mechanics, this paper will seek to establish formal models applied to mental information and then to test them experimentally using a Remote State Preparation of Mental Information protocol.

The mathematical formalism of quantum information theory can be applied for description of any kind of statistical data which cannot be embedded into the Kolmogorov model. Therefore the domain of possible applications of quantum information theory is not reduced to quantum physics. It might be successfully applied in a variety of research areas, *e.g.,* cognitive science and psychology (Khrennikov, 2004). In this paper we present one of such applications. The hypothesis that the human mind can reveal quantum-like properties is a recent field of investigation, currently approached from both a theoretical and an experimental point of view. This line of research is part of the interdisciplinary research known as *quantum cognition*, whose main objective is to verify whether it is possible to generalize for mental (and biological) variables what is expected theoretically and experimentally verified with physical variables. More precisely, this paper aims to *study the application of a quantum mechanics protocol to cognitive science*.

It was necessary to wait about 55 years (circa 1925-1980) to obtain the first experimental data supporting the theoretical models underlying quantum mechanics, in particular the possibility of quantum non-locality related to entanglement between physical objects. Since the 80's, there has been an exponential increase of experimental evidence supporting the theoretical models, see (Genovese, 2005; 2010) for a review. Among the most interesting developments is the observation that these phenomena are observed in physical objects in ever greater complexity and at increasingly high temperatures, contradicting the hypothesis that these phenomena may be confined to mere atomic/subatomic physical objects such as photons, atoms, etc. and at temperatures incompatible with biological metabolism (Vedral, 2010). In parallel with this generalization of the laws of quantum mechanics to physical objects, an interest has grown in modeling and in experimentally testing whether methods and phenomena observed in quantum mechanics could also find application in human sciences concerned with both mental and biological variables (Khrennikov, 2010; Asano *et al.*, 2011). With regards to mental variables, experimental studies have already been applied to cognitive processes such as reasoning and the organization of the lexicon (Aerts, 2009; Busemeyer *et al.*, 2011) supporting the validity of the hypothesis that mental processes can express quantum-like characteristics. This paper fits into this line of research. It aims to define theoretically and experimental testing, the possibility that human cognition can manifest phenomena of entanglement-like correlations at a distance, apparently violating the laws of classical physics, which state that information can only be connected through electromagnetic signals conveyed by the sensory organs.

Among the different protocols used to test entanglement in QM, we discovered the possibility of extending the Remote State Preparation protocol used for quantum information, to a mental one, which we renamed Remote State Preparation of Mental Information (RSPMI). The Remote State Preparation protocol allows preparing at distance, which is without using the classical channel of communication, the status of physical observables (*i.e.,* photons, atoms, etc.) sent by Alice the sender, to Bob, the receiver, exploiting their entangled status. The basic RSPMI protocol, strictly maintains the characteristics of that used in physics, only changing the nature of Alice and Bob, from electronic devices to real human beings and the type of information, from photons to visual images. Alice, the sender, has full knowledge of the information she wants to send to Bob. Bob, the receiver, has zero knowledge of the information that Alice wants to send him a part from the fact that they are pictures. The paper therefore aims: a) to define a mathematical model of the limits of communication based on the expected probability with classical communication resources, and b) to verify experimentally whether those boundaries are violated, suggesting the possibility of communication using a state of entanglement-like correlation between Alice and Bob.

### 1.2. *Remote State Preparation*

In Quantum Mechanics (QM) Remote State Preparation (RSP), is a variant of teleportation where Alice has full knowledge of the state she intends to prepare at Bob's location. This protocol was first proposed by Pati (2000) and generalised for arbitrary *q*bits by Bennet et al. (2001).

The goal of RSP is to prepare a quantum state at a distant location, without sending the actual state. Alice, the sending party, knows exactly the target state $\rho tar$ that she wants Bob, the receiving party, to have.

Several features are usually desired in an RSP protocol: Bob should need limited or zero knowledge of the state Alice is trying to prepare, and the required communication resources (classical and/or quantum) should be limited. Perhaps most importantly, the protocol should yield output states ρ*out* at Bob's location which closely match the target states ρ*tar* which Alice intended to prepare.

However, due to the practical limitations of imperfect devices, no RSP experiment can yield remotely prepared output states which exactly match the intended states. Indeed, we should be satisfied when the output states have a high fidelity with the intended states. This raises the question: how high must this fidelity be, on average, for an experiment to demonstrate a genuine quantum advantage? In other words, if we restrict Alice and Bob to a comparable, fixed amount of classical communication—but no shared entanglement—what is the optimal average RSP fidelity they could achieve?

It is only when an experiment surpasses such a classical threshold that we can be sure of having demonstrated verifiable advantages to quantum communication.

Ideally, the fidelity should be *F*(ρ*tar*, ρ*out*) = 1 for any target state.

Unlike teleportation, Alice accesses the state index, not the state, though she has complete information about the state and may prepare herself a copy if desired. She communicates a message to Bob, sending a limited number *c* of classical bits (*c*bits). Bob then prepares an output state ρ*out*. Their goal is for the output states to match the target states with the highest possible quantum fidelity, on average, that is, to maximize the quantity.

### 1.3 *RSP Classical probability boundaries*

Killoran *et al.,* (2010) studied the fidelity achievable by RSP protocols lacking shared entanglement and determine the optimal value for the average fidelity in several different cases. In the classical case, where Alice and Bob share no entanglement, to find the optimal achievable fidelity, it is assumed that Alice and Bob are unencumbered by the imperfections of real-world devices.

Considering the situation where the target ensemble consists of a finite number of states, we may assume that we have fixed a *finite* ensemble of target states each with a fixed probability

$p\alpha \{\rho^{tar}\alpha, p\alpha\}^n{}_{\alpha=1}$.

Alice sends a string of c *c*bits. We can label all messages of this type by a natural number $m(\alpha) = k \in \{0, 1, \ldots, 2c - 1\}$. In general, the message assignment may be either deterministic [e.g., $m(\alpha) = 3$] or probabilistic; that is, $m(\alpha) = k$ with probability $qk(\alpha)$, where for each $\alpha$, $\Sigma k\, qk(\alpha) = 1$.

Upon receiving the message $k$, Bob prepares an output state $\rho_{out}$ aiming to reproduce Alice's $\rho^{tar}{}_\alpha$. Their goal is for the output states to match the target states with the highest possible quantum fidelity, on average, that is, to maximize the quantity $F = \Sigma_\alpha\, (p_\alpha F(\rho^{tar}{}_\alpha, \rho^{out}{}_\alpha)$.

Here follows the algorithm devised by Killoran *et al.*, (2010) for determining the classical probability upper bounds will be outlined.

### 1.4 RSP threshold calculations

Consider a fixed a finite ensemble of target states $\{\rho tar\, \alpha, p\alpha\}n\, \alpha=1$. It is clear that whenever $n \leq 2^c$, the optimal classical protocol can achieve perfect fidelity since there is sufficient capacity in the message to uniquely label the state. The interesting cases have $n > 2^c$. The optimum average fidelity can be determined by checking the value of Eq. (1) for all partitionings of the $n$ target states into $2^c$ disjoint subsets, but this can be inefficient even for modest values of $n$ and $c$. Alternatively, we search for an upper bound on the threshold which is easier to calculate. If an experiment surpasses the upper bound, it has surpassed the actual threshold.

Killoran *et al.*, (2010) defined an efficient algorithm for determining such upper bounds. For this algorithm, they make the additional assumption that each target state has equal probability to be chosen from the target ensemble. They note that each partition contains some number $s$ of states and contributes one term to the sum in Eq.(1).

$$\langle F \rangle \max = \sum_{k=0}^{2^c-1} pk\lambda_k^{max} \quad (1)$$

Two different partitions with the same number of states may contribute differently to the average fidelity, depending on the arrangement of the states. However, for each number $s \in \{0, 1, \ldots, 2^c - 1\}$, there is a set of $s$ states which yields the maximal possible contribution $\langle F \rangle_s^{max}$.

By using these maximal values in Eq. (1) instead of the actual values, they obtained an upper bound on the threshold.

The first step in the algorithm involves checking all partitions of size $s$ to find the maximal contribution $\langle F \rangle_s^{max}$

Next, they listed all the ways in which $n$ elements can be divided into $2^c$ subsets. The order of the subsets does not matter, so for simplicity it is possible to create a list in order of decreasing partition size. This list forms a table with $2^c$ columns. For each row $i$, a list of numbers $\{sij\}^{2^c-1}{}_{j=0}$

which sum to $n$. To determine the upper bound, they calculated the quantity (Eq. 2)

$$\langle F \rangle i = \sum_{k=0}^{2^c-1} \frac{sij}{n} \langle F \rangle_{sij}^{max} \quad (2)$$

The highest $\langle F \rangle_i$ provide an upper bound on the optimal average fidelity.

It may even be the case that the threshold is equal to the upper bound found via the above algorithm, especially if the target ensemble exhibits a high degree of symmetry. To verify this, one would have to find a specific partitioning which leads to the same value as the upper bound. On the other hand, if it can showed through other arguments, that the highest $\langle F \rangle_i$ is unachievable, then the second-highest $\langle F \rangle_i$ provides a new, smaller upper bound.

The experimental evidences presented by Killoran et al. (2010), confirm the violation of this limit using the RSP protocol. For all nonunity benchmarks, the experimental values surpass the benchmarks for two (three) transmitted *c*bits by at least 46 times the standard error of the mean.

Even if this communication protocol has been exclusively considered within the communication of quantum states, in principle it may be applied in the communication of any type of information, biological or mental.

## 2. Discussion on measurements on composite physical and mental systems

A composite (physical) quantum system S of, *e.g.,* two electrons, $S\_1$, $S\_2$, in the entangled state cannot be more considered as a pair of two systems ($S\_1$, $S\_2$). This is a single system S. Any measurement on S is measurement on the whole system. The most interesting are measurements on S performed by two observables say A and B at two different spatial locations, but at matching instances of time $t\_1=t\_2$. Each observable, either A or B, is an observable on the whole system. This holistic structure of a composite quantum system can be exhibited in the form of "quantum nonlocality".

In the mental setting, the system of uniformly distributed ensemble of four pictures can be considered as a quantum-like system S. We emphasize that this system is *purely informational*, see Khrennikov (2004) on a model of purely informational reality unifying both physical and mental phenomena. There are two observables on this system Alice and Bob, the RSPMI experiment can be considered as a joint measurement performed by a pair of observables A (Alice) and B (Bob). The violation of Bell's inequality is a sign of quantum-like holistic effects for mental informational states. The version of RSPMI experiment, in which a computer plays the role of Alice (see RSPMI variants paragraph 4.2.1), can be also embedded in the aforementioned scheme, the role of the A-observables is played by the physical random generator and computer.

We remark that the *informational interpretation* of conventional quantum mechanics plays an important role in justification of our purely informational model of mental entanglement and the RSPMI experiment. The idea that quantum theory is not about particles nor waves, but about information and the latter is the fundamental element of quantum reality was discussed in works of leading experts in quantum foundations, e.g., Zeilinger (2010) and Fuchs (2002). Of course, these authors wrote about information obtained from physical systems, but the usage of this interpretation for cognitive systems (Khrennikov, 2004) is quite natural.

## 3. The RSMI standard protocol

We now outline a RSP protocol dealing with the communication of mental information (RSPMI) determining the boundaries of a communication based on the probability expected with classical resources to see whether these boundaries are violated. In this case, this finding will suggest the possibility of communication exploiting an entanglement state between Alice and Bob.

The proposal of a RSPMI protocol is new but the possibility of non-local communication of mental information has been heavily investigated and is one of the main aims of quantum cognitive psychology. Within this discipline, the hypothesis that the human mind may express non local or quantum-like characteristics has a long history starting before Aspect's first experimental demonstrations of quantum entanglement (Aspect *et al.,* 1982a; 1982b) and continuing up recently with the seminal books and papers of Radin (2006), von Lucadou, Romer and Walach (2007) and Khrennikov (2010) even if this interpretation is not shared by all researchers.

The basic protocol of RSPMI strictly maintains the characteristics of that used in physics changing only the nature of Alice and Bob, from electronic devices to real human individuals and the information, from photons to visual pictures. Alice, the sender, has full knowledge of the state of the information that she wants to send to Bob. To support Alice's choice, each target state is usually chosen by a built-in RND pseudo-algorithm or real RNG connected to a computer.

Our protocol has a finite ensemble of four target states $p\alpha = 1/4$. The four states have a uniform probability distribution and ¼ probabilities to be chosen by Alice.

Usually there are four different pictures or four different short ($\approx$ one minute) video clips. There is no agreed consensus about the criteria to choose them apart from the suggestion that they be maximally differentiated both from a semantic content and visual characteristics to facilitate Bob's discrimination.

Bob, the receiver, has zero knowledge of the information Alice wants to send him apart from the fact that it is a picture. His task is to reproduce or identify Alice's information. Coincidences correspond to the Bob's recognition of Alice targets by drawing or sketching the picture, by describing it verbally, or by identifying it within a sample, usually a set of four different pictures. A common alternative to the identification of Alice's information is its identification by independent judges comparing Bob verbal or written description with a set of four different pictures. In a typical experiment, many Alices (participants) are recruited to send a state (target) to each Bob (different participants), usually known to Alice. Usually, the number of states sent by Alice may vary from one to ten. In the RSPMI basic protocol we assume that entanglement between Alice and Bob is obtained when the two parties met and agreed on the protocol procedure. The shared mental representation of the correspondent part assures a mental entanglement, even when they are at distance without any possibility of classical communication.

The result of each experiment is the average percentage of correct coincidence (hits) among all Alices and Bobs detected by electronic mean or by independent judges. The main similarities and differences between the standard RSP and the RSPMI protocol are presented in Table 1 and sketched in Figure 1

**Table 1.** Similarities and differences between the standard RSP and RSPMI protocol.

|  | RSP | RSPMI |
|---|---|---|
| **Alice identity** | Electronic device | Electronic device or Human being |
| **Bob identity** | Electronic device | Human being |
| **Alice initial knowledge** | Target complete knowledge | Target complete knowledge |
| **Bob initial knowledge** | Target zero knowledge | Target zero knowledge |
| **Information type** | *q*bits | Classical (i.e. images, video clips) |
| **Entanglement mode** | i.e. Parametric downconversion | Mental connection |
| **Transmitted *c*bits** | 2(3) | 0 |
| **Locality loophole** | Partially closed | Closed for sensory information |
| **Fair-sampling loophole** | open | closed |
| **Events per experiment** | thousands | Usually less than one hundred |
| **Time per event** | Fraction of seconds | 15 to 30 minutes |
| **Coincidence counts** | Electronic device | Electronic device or independent judge |

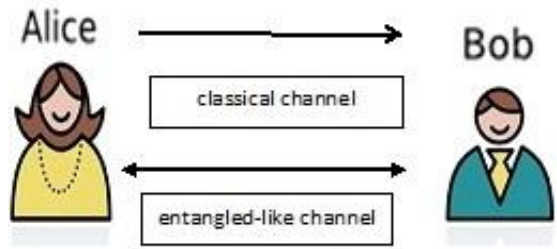

**Figure 1.** Sketch drawing of Alice and Bob in the RSPMI protocol.

With a RSPMI protocol and some of its variants (see paragraph RSPMI variant), the *locality loophole* arising when Alice's measurement result can in principle be causally influenced by a physical (subluminal or luminal) signal from Bob's measurement event or Bob's choice event, and vice versa, is closed only for classical communication. The best available way to close this loophole is to space-like separate every measurement event from both the measurement, outcome independence on one side and setting choice on the other side (Jarrett, 1984). In a RSPMI protocol, Alice and Bob are always spatially separated by some meters and any possibility of sensory communication between the two, or to receive Alice's information by other human or technical means is eliminated, even if the distance cannot exclude a subluminal or luminal communication. The second loophole, the *fair-sampling loophole* (Pearle, 1970; Adenier *et al.,* 2007), arises from inefficient particle collection and detection. It suggests that, if only a fraction of generated particles is observed, it may not be a representative subensemble, and an observed violation of Bell's inequality could still be explained by local realism, with the full ensemble still obeying Bell's inequality. In a standard RSPMI protocol, every unit of information sent by Alice is received by Bob and none unit of information is lost. Some information unit may be discarded in the data analysis but only in case of a clear violation of the protocol or error in the data analysis.

## Material and Methods
### 4.1 *RSPMI benchmarks threshold calculations*
Given the characteristics of RSPMI protocol, we restrict ourselves to ensembles of pure deterministic states with a uniform distribution: $p\alpha = 1/n$. Our goal is to find benchmarks to be experimentally surpassed. If this classical limit gets violated we can assume that Alice and Bob share an initial supply of maximally entangled information.

If the theoretical $F(\rho tar, \rho out) = 1$ for any target state, we can derive the expected F value if Alice and Bob do not share any entangled information. Given the "imperfections" of the measurement apparatuses, we can accept, $F = 0.9$ as optimal. If as benchmark we assume the fidelity expected by

$p\alpha = ¼$, the resulting fidelity value is $F= 0.7482029$ (see calculation details in the Appendix) representing the limit to be violated if Alice and Bob share entangled information.

### 4.2 *Experimental evidences*
A summary of all experiments respecting the described standard RSPMI protocol carried out up 2011[2], using states with a uniform distribution: $p\alpha = 1/4$ with Bob in a special mental condition named ganzfeld (Wackermann *et al.,* 2008), devised to reduce "mental noise", to facilitate target identification, includes 87 experiments, for a total of 3338 events (trials) and an average of 33.82 % of coincidences weighting the result of each study for √number of trials. This database is considered homogeneous even if experiments have been carried out by different experimenters and different participants.

---

[2] The complete list of references is available upon request to the first author.

The overall experimentally achieved mean fidelity is $F = 0.808969964 \pm 0.001463$ violating the benchmark $F = 0.7482029$ of 41.5 standard units.

### 4.2.1 RSPMI variant
Among the most used variants, Alice is substituted with a computer that randomly picks up one out four pictures of a set. This choice is usually performed before Bob's target identification.

In this variant, it is assumed that the entanglement is established between Bob and the distant information even if represented in an electronic device and not in a human mind as in the basic RSPMI protocol (see paragraph 3).

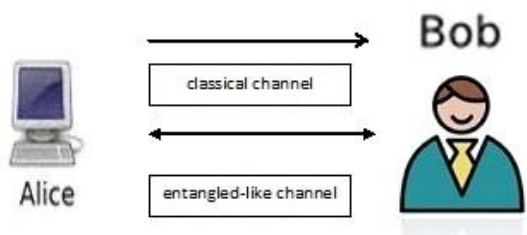

**Figure 2.** Sketch drawing of Alice and Bob in the RSPMI protocol variant.

The overall experimentally achieved mean weighted fidelity observed in 24 experiments with a total of 1275 trials and an weighted average coincidences of 28.4%, $F = 0.773165374 \pm 0.00061946$ violating the expected $F = 0.7482029$ of 40.29 standard units.

## 5. Conclusion
The theoretical and the summary of experimental evidence currently available, seems to support the possibility of RSPMI. The basic protocols where both Alice and Bob are human with Bob in a special mental condition to reduce "mental noise" in order to enhance the noise-signal ratio, the Fidelity benchmark of a classical connection is violated by more than 45 standard units. A variant of the basic protocol where Alice is a computer, the violation from the expected classical information communication, is violated by more than 40 standard units.

Even if the RSP and RSPMI protocols share many characteristics, in particular those related to Alice and Bob initial knowledge much has to be studied to know how far the RSP and RSPMI analogy hold up. It is evident that the transmission of information between two human minds and two electronic devices are absolutely different in terms of efficiency, noise-signal ratio, measurement and so on. Furthermore another basic difference is related to the type of information shared between Alice and Bob. Physical information like photons, atoms, etc., is absolutely different from mental ones and how it is processed (measured).

Furthermore, whereas RSP benchmarks for classical communication is calculated referring to the number of $c$bits Alice send to Bob, in the RSPMI protocol, the classical benchmark is derived from the $p\alpha = 1/n$.

Another important difference is related to the characteristics of entanglement. Whereas at present it is quite easy to entangle two or more physical objects, the "entanglement" between human minds or between a human mind and information at distance is assumed from the theoretical model presented in the paragraph 4 (see RSP and RSPMI comparison in Table 1).

However, the experimental evidence accumulated up until today supports the feasibility of a RSPMI and there is good evidence of a genuine entanglement communication at distance with two RSPMI protocols.

The available experimental evidence supports the fact that it is possible to unify physical and mental objects in terms of units of information (Khrennikov, 2004), bringing a change to the concept of information and promoting the integration between disciplines that have as their focus different aspects of reality, the physical and mental, leading to a significant paradigm shift in cognitive and information science.

## Acknowledgments
We thank Marco Genovese and Ivano Ruo Berchera for their contributions and suggestions in particular for the data analysis.

## Appendix

Calculation details related to the fidelity estimation and the comparison with the benchmark and experimental results.

### General equations:

### Fidelity:

$F = \Sigma_{(i)} \sqrt{(p(i) * q(i)} + \sqrt{(1- p(i)) * (1-q(i))}$, where p(i), the theoretical probability and q(i) the experimental probability.

Propagation of the uncertainity of the mean:

$SE/2 * |\sqrt{(p(i) * q(i)} - \sqrt{(1- p(i)) * (1-q(i))}|$, where SE = standard error of mean.

Benchmark:

$F = \sqrt{(0.9 * 0.25)} + \sqrt{(0.1 * 1-0.25)} = 0.7482029$

Experimental fidelity with the classical RSPMI protocol:

$F = \sqrt{(0.9 * 0.338214)} + \sqrt{(0.1 * 1-0.338214)} = 0.808969964 \pm 0.00146323$

Experimental fidelity with the RSPMI protocol variant:

$F = \sqrt{(0.9 * 0.284)} + \sqrt{(0.1 * 1-0.284)} = 0.773165374 \pm 0.000619459$